\documentclass[epjST]{svjour}
\usepackage{graphics}
\usepackage{mathtools}
\usepackage{amssymb}
\usepackage{amsmath}
\usepackage{xcolor}
\usepackage[numbers,sort&compress]{natbib}

\newcommand{\moy}[1]{\left\langle #1 \right\rangle}

\newcommand{\XX}[0]{\boldsymbol{X}}

\newcommand{\ex}[1]{\mathrm{e}^{#1}}

\newcommand{\dd}[0]{\mathrm{d}}

\newcommand{\RR}[0]{\boldsymbol{R}}

\newcommand{\kB}[0]{k_{\mathrm{B}}}
\newcommand{\LL}[0]{\boldsymbol{l}}
\newcommand{\nn}[0]{\hat{\boldsymbol{n}}}

\newcommand{\MM}[0]{\mathbf{M}}
\newcommand{\WW}[0]{\mathbf{W}}
\newcommand{\GG}[0]{\mathbf{\Gamma}}

		\newcommand{\muO}[0]{\boldsymbol{\mu}^{11}}
		\newcommand{\muT}[0]{\boldsymbol{\mu}^{22}}
		\newcommand{\muOT}[0]{\boldsymbol{\mu}^{12}}

		\newcommand{\II}[0]{\boldsymbol{1}}
		\newcommand{\sgm}[0]{\boldsymbol{\sigma}}
		
		\newcommand{\Vpho}[0]{\mathbf{\Lambda}^v}
		\newcommand{\vpho}[0]{\mathbf{\Lambda}^\omega}

\begin{document}
\title{Diffusion and steady state distributions of flexible chemotactic enzymes}
\author{Jaime Agudo-Canalejo\inst{1,2,3}\fnmsep\thanks{\email{jaime.agudo@ds.mpg.de}} \and Ramin Golestanian\inst{1,2}\fnmsep\thanks{\email{ramin.golestanian@ds.mpg.de}}}
\institute{Max Planck Institute for Dynamics and Self-Organization (MPIDS), D-37077 G\"ottingen, Germany \and Rudolf Peierls Centre for Theoretical Physics, University of Oxford, Oxford OX1 3PU, United Kingdom \and Department of Chemistry, The Pennsylvania State University, University Park, Pennsylvania 16802, USA}
\abstract{
Many experiments in recent years have reported that, when exposed to their corresponding substrate, catalytic enzymes undergo enhanced diffusion  as well as chemotaxis (biased motion in the direction of a substrate gradient). Among other possible mechanisms, in a number of recent works we have explored several \emph{passive} mechanisms for enhanced diffusion and chemotaxis, in the sense that they require only binding and unbinding of the enzyme to the substrate rather than the catalytic reaction itself. These mechanisms rely on conformational changes of the enzyme due to binding, as well as on phoresis due to non-contact interactions between enzyme and substrate. Here, after reviewing and generalizing our previous findings, we extend them in two different ways. In the case of enhanced diffusion, we show that an exact result for the long-time diffusion coefficient of the enzyme can be obtained using generalized Taylor dispersion theory, which results in much simpler and transparent analytical expressions for the diffusion enhancement. In the case of chemotaxis, we show that the competition between phoresis and binding-induced changes in diffusion results in non-trivial steady state distributions for the enzyme, which can either accumulate in or be depleted from regions with a specific substrate concentration.
} 
\maketitle
\section{Introduction}
\label{intro}

Enzymes have attracted much attention in recent years as biocompatible nanomachines that may perform work and undergo directed motion, with many biomedical and nanoengineering applications \cite{gole05,seng14b,dey15,ma16}. In particular, much work has been devoted to understanding and further exploring experimental observations of enhanced diffusion \cite{mudd10,seng13,ried15,mikh15,gole15,bai15,illi17a,illi17b,jee17,jee18,gunt18,zhan18,feng19,gunt19,xu19,jee19,adel19a,hosa19} and chemotaxis \cite{yu09,seng13,dey14,seng14,weis17,agud18a,zhao17,jee17,jee18,moha18,adel19b} of enzymes in the presence of their substrate. Chemotaxis, in particular, may have important implications in the self-organization of enzymes that participate in a common catalytic pathway \cite{wu15,zhao17,swee18,agud19}.

The mechanism, or more appropriately mechanisms, underlying enhanced diffusion and chemotaxis are still far from being fully understood \cite{agud18c}. Early attempts at an explanation focused on active mechanisms, involving the non-equilibrium activity arising from the (typically exothermic) catalytic step in which the enzyme transforms a substrate molecule into a product molecule \cite{ried15,mikh15,bai15}. However, a systematic investigation \cite{gole15} of several active mechanisms, including self-phoresis \cite{gole05}, reaction-induced boost in kinetic energy \cite{ried15}, stochastic swimming \cite{Golestanian:2008b,Golestanian2009b,Golestanian2010,Sakaue:2010,Najafi2010,bai15,gole15}, and collective heating of the enzyme \cite{gole15}, showed that none of them is strong enough to account for the observed values of diffusion enhancement, which range from about $\sim$20\% to as high as 200\% \cite{xu19}. Recent experimental observations of ballistic motion for urease and acetylcholinesterase \cite{jee17,jee18}, however, do seem to suggest the existence of an active mechanism, but it is still unclear what kind of effect could account for such directed motion \cite{gole05,feng19}.

The observation of enhanced diffusion for aldolase, a slow and endothermic enzyme, which moreover was observed not only in the presence of its substrate but also in the presence of an inhibitor (which binds to the enzyme but does not induce a catalytic step) \cite{illi17a}, demanded a change of paradigm from active to passive mechanisms. In this context, we have shown that conformational changes of the enzyme induced by specific binding to the substrate (or inhibitor) may be sufficient to account for enhanced diffusion \cite{illi17b,adel19a}. These conformational changes include not only changes in the average shape of the enzyme, but also changes in its shape fluctuations. Furthermore, we have shown that, in the presence of a substrate gradient, binding-induced conformational changes and phoresis compete against each other and pull the enzyme in opposite directions \cite{agud18a}, a mechanism which may explain conflicting observations in the direction of urease chemotaxis \cite{seng13,jee17}. Anisotropic enzymes may also undergo alignment in the presence of gradients of substrate or of the enzyme itself \cite{adel19b}.

In this paper, we will extend and refine these results in several ways. In Section~\ref{methods}, we recapitulate our previous work \cite{illi17b,agud18a,adel19a,adel19b}, and generalize it to an arbitrary choice of tracking point on the enzyme for which the overall translational diffusion and drift of the enzyme are calculated. In Section~\ref{taylor}, we use generalized Taylor dispersion theory \cite{bren82,bren87,bren93} to obtain exact and explicit expressions for the long-time diffusion coefficient and binding-induced changes of enzyme diffusion. We also show that the long-time diffusion coefficient is independent of the choice of tracking point, and is always smaller than the short-time diffusion coefficient, which does depend on the choice of tracking point.  Lastly, in Section~\ref{steady} we calculate the steady state distribution of enzymes in the presence of a substrate gradient, and show that the competition between phoresis and binding-induced conformational changes may cause accumulation or depletion of the enzyme from specific regions in space.

\section{Diffusion-drift of a dumbbell enzyme in the presence of a substrate gradient} \label{methods}

\subsection{Theory and closure approximation}

\begin{figure}
\centering
\includegraphics[width=1\linewidth]{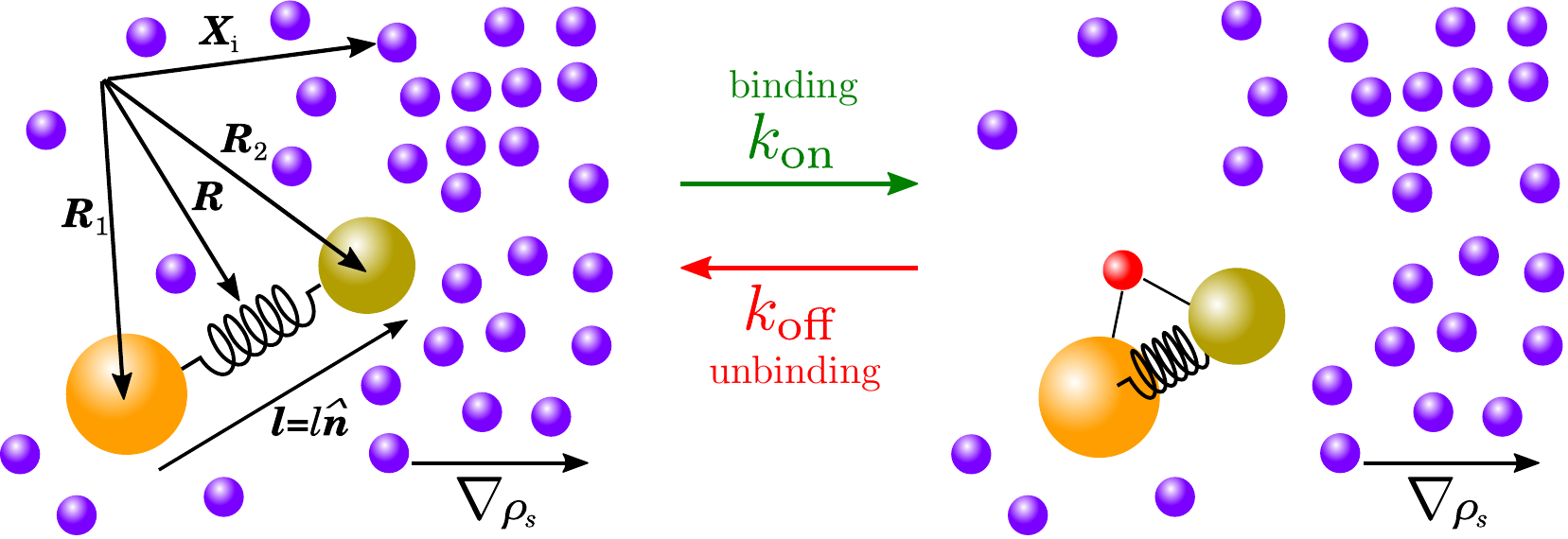}
\caption{We use a flexible, asymmetric dumbbell as a minimal model for an enzyme which is composed of two subunits and can undergo conformational fluctuations. The two subunits interact with the substrate \emph{via} non-contact interactions (van der Waals, electrostatic...) and hydrodynamic interactions. Moreover, the substrate molecules can bind specifically to the enzyme, in which case the enzyme may undergo conformational changes that affect its average length or its rigidity.}
   \label{fig:dumbbell}
\end{figure}

As a minimal model of an enzyme with internal structure and anisotropic shape, let us consider a flexible dumbbell with two spherical subunits, which may have different sizes and surface properties, in a gradient of solute particles, see Fig.~\ref{fig:dumbbell}. The location of the dumbbell subunits is denoted by $\RR_1$ and $\RR_2$, and the location of the substrate particles by $\XX_i$ with $i=1,\dots,N$. The full $N+2$-particle distribution for the dumbbell in contact with the substrate particles is
\begin{eqnarray}
&& \partial_t\rho_{N+2}(\RR_1,\RR_2,\XX_1,\dots,\XX_N;t)=\nonumber\\
&&  \sum_{i,j=1}^2 \nabla_{\RR_i} \cdot \left( \boldsymbol{\mu}^{ij}\cdot[\kB T \nabla_{\RR_j}\rho_{N+2}+  (\nabla_{\RR_j}\phi^{{N+2}})\rho_{N+2}] \right) +\sum_{i=1}^N \Big\{ \sum_{j=1}^2  \Big[ \nabla_{\RR_j} \cdot \left( \boldsymbol{\mu}^{js}\cdot[\kB T \nabla_{\XX_i}\rho_{N+2}+ (\nabla_{\XX_i}\phi^{{N+2}})\rho_{N+2}] \right) \nonumber\\
&&+ \nabla_{\XX_i} \cdot \left( \boldsymbol{\mu}^{sj}\cdot[\kB T \nabla_{\RR_j}\rho_{N+2}+  (\nabla_{\RR_j}\phi^{{N+2}})\rho_{N+2}] \right) \Big]+\nabla_{\XX_i} \cdot \left( \boldsymbol{\mu}^{ss}\cdot[\kB T \nabla_{\XX_i}\rho_{N+2}+ (\nabla_{\XX_i}\phi^{{N+2}})\rho_{N+2}] \right) \Big\}
\end{eqnarray}
where $\boldsymbol{\mu}^{ij}$ with $i,j=1,2,s$ are hydrodynamic mobilities, and the interaction potential for the full system is
\begin{eqnarray}
\phi^{{N+2}}(\RR_1,\RR_2,\XX_1,\dots,\XX_N) = U(\RR_1-\RR_2) + \sum_{j=1}^2 \sum_{i=1}^N \phi^{jB}(\RR_j-\XX_i)
\end{eqnarray}
where $U$ is the pair potential between the two dumbbell subunits, and $\phi^{js}$ is the pair potential between subunit $j$ and a substrate particle.

We can define the two-particle distribution describing the dumbbell, by integrating out all the degrees of freedom corresponding to the substrate particles
\begin{eqnarray}
\rho_{12}(\RR_1,\RR_2) & = & \int \dd \XX_1 \dots \dd \XX_N\,  \rho_{{N+2}}(\RR_1,\RR_2,\XX_1,\dots,\XX_N)
\end{eqnarray}
Taking this integral in the equation for the full distribution above, one would obtain an equation for the two-particle distribution that depends on the three-particle distribution for the two dumbbell subunits and a substrate molecule $\rho_{12s}(\RR_1,\RR_2,\XX)$. The equation for the three-particle distribution would in turn depend on the four-particle distribution, and so on. In Ref.~\cite{adel19b}, we showed how this infinite hierarchy of equations can be truncated using the closure approximation
\begin{equation}
\rho_{12s}(\RR_1,\RR_2,\XX) \simeq \rho_{12}(\RR_1,\RR_2) \frac{\rho_\mathrm{s}(\XX)}{N} \ex{-\frac{\phi^{1s}+\phi^{2s}}{\kB T}}
\label{eq:dbclosure}
\end{equation}
where $\rho_\mathrm{s}$ represents the externally imposed substrate concentration profile. Using the closure we obtain a closed evolution equation for the two-particle distribution
\begin{eqnarray}
\partial_t \rho_{12}(\RR_1,\RR_2;t)  = &&    \sum_{i,j=1}^2 \nabla_{\RR_i} \cdot \left\{ \boldsymbol{\mu}^{ij} \cdot  \left[ \kB T \nabla_{\RR_j}\rho_{12} + (\nabla_{\RR_j} U)\rho_{12}   \right] \right\}  \label{eq:twopart3} \\ && -  \nabla_{\RR_1} \cdot \left\{  \rho_{12} \sgm_1 \cdot \nabla_{\RR_1} \rho_\mathrm{s} \right\} -  \nabla_{\RR_2} \cdot \left\{  \rho_{12} \sgm_2 \cdot \nabla_{\RR_2} \rho_\mathrm{s} \right\} \nonumber
\end{eqnarray}
with
\begin{equation}
\sgm_1 \equiv  \frac{\kB T}{\eta} \left[A_1 + \frac{a_2^3}{l^3} \left( B_2 - \frac{3}{2} A_1 \right) \left(  \nn \nn - \frac{\boldsymbol{1}}{3} \right) \right]
\end{equation}
\begin{equation}
\sgm_2 \equiv  \frac{\kB T}{\eta} \left[A_2 + \frac{a_1^3}{l^3} \left( B_1 - \frac{3}{2} A_2 \right) \left(  \nn \nn - \frac{\boldsymbol{1}}{3} \right) \right]
\end{equation}
corresponding to the phoretic mobilities of subunits 1 and 2, respectively. The phoretic mobilities arise from the interactions between the subunits and the substrate \emph{via} the potentials $\phi^{js}$. The first term in the phoretic mobilities is due to the individual response of each subunit to the gradient, with the coefficient $A_i$ given by
\begin{equation}
A_i \equiv \frac{1}{6 a_i} \int^\infty_{a_i}\text{d}r_i\,r^2_i\,(\ex{-\frac{\phi^{is}}{\kB T}}-1)\left(4-4\frac{a_i}{r_i}+\frac{a_i^4}{r_i^4}-\frac{a_i^6}{r_i^6}\right)
\label{Aig}
\end{equation}
where $r_i$ is the distance from the center of subunit $i$. For particles that are much larger than the range of the interaction, we can use $r = a_i + \delta$ with $\delta \ll a_i$. The terms inside the rightmost parenthesis in the integral become $6 \delta/a_i$ to lowest order, giving
\begin{equation}
A_i \approx \int_0^\infty \mathrm{d}\delta \delta ( \ex{- \phi^{is}(\delta)/\kB T} - 1) \equiv  \lambda_i^2
\label{Ai}
\end{equation}
which shows that the coefficient $A_i$ is a generalization of the usual Derjaguin length $\lambda_i$ used to describe phoresis \cite{derj47,gole09,agud18a}, to the case in which the particle is not necessarily larger than the interaction range. The second term in the phoretic mobilities, proportional to $a_i^3/l^3$ where $l$ is the distance between the centers of the two subunits, represents corrections to the phoretic mobility due to the presence of the nearby subunit. The coefficients $B_i$ are given by
\begin{equation}
B_i \equiv \frac{1}{10} \int^\infty_{a_i}\text{d}r_i\,r_i\,(\ex{-\frac{\phi^{is}}{\kB T}}-1) \left( 1 - 5 \frac{r_i}{a_i} + 5 \frac{r_i^3}{a_i^3} \right)
\label{Big}
\end{equation}
which, considering very short ranged interactions when $r = a_i + \delta$ with $\delta \ll a_i$, becomes
\begin{equation}
B_i \approx \frac{a_i}{10} \int^\infty_0 \text{d}\delta\,(\ex{-\frac{\phi^{is}}{\kB T}}-1) \equiv \frac{a_i}{10} \gamma_i
\label{Bi}
\end{equation}
where we have defined $\gamma_i$, which is a lengthscale of the order of the interaction range, but distinct from the Derjaguin length $\lambda_i$. 

\subsection{Separation into position and internal degrees of freedom}

Because we are mainly interested in the long time diffusion and drift of the dumbbell enzyme as a whole, ignoring its internal degrees of freedom (elongation and orientation of the dumbbell), it is convenient to transform of Eq.~\ref{eq:twopart3} to coordinates representing the overall position of the dumbbell, and the state of its internal degrees of freedom. The internal degrees of freedom are most approprately represented by the elongation vector $\LL \equiv \RR_2 - \RR_1$, which in turn can be decomposed into an elongation scalar $l$ and a unit orientation vector $\nn$, so that $\LL = l \nn$. To identify the location of the dumbbell, we choose an arbitrary tracking point along the line connecting the two subunits, given by  $\RR = \RR_1 +  g(l) \LL$, where $g(l)$ is an arbitrary function of the elongation. For example, the choice $g(l)=1/2$, which was used in our previous works \cite{illi17b,adel19a,adel19b}, would correspond to tracking the midpoint between the two subunits. Some subtleties related to the choice of tracking point will be discussed in Section~\ref{taylor}.

With such a general choice of tracking point as determined by the choice of $g(l)$, the diffusion-drift equation (\ref{eq:twopart3}) can be written as
\begin{eqnarray}
\partial_t\rho_{12}(\RR, \boldsymbol{l}; t) &= & \nabla_{\RR}\cdot(\MM\cdot\kB T\nabla_{\RR}\rho_{12}) + \nabla_{\boldsymbol{l}}\cdot(\GG\cdot\kB T\nabla_{\RR}\rho_{12}) \nonumber \\ &+& \nabla_{\RR}\cdot\left[\GG\cdot(\kB T\nabla_{\boldsymbol{l}}\rho_{12}+(\nabla_{\boldsymbol{l}}U)\rho_{12})\right] + \nabla_{\boldsymbol{l}}\cdot\left[\WW\cdot(\kB T\nabla_{\boldsymbol{l}}\rho_{12}+(\nabla_{\boldsymbol{l}}U)\rho_{12})\right] \nonumber \\
&-&  \nabla_{\RR} \cdot [ \rho_{12}  \Vpho \nabla_{\RR} \rho_\mathrm{s} ] - \nabla_{\LL} \cdot [ \rho_{12} \vpho  \nabla_{\RR} \rho_\mathrm{s} ] \label{smol1}
\end{eqnarray}
with the translation tensor
\begin{equation}
\MM = M_I \II + M_D \nn \nn
\label{Mgen}
\end{equation}
with components
\begin{equation}
M_I = (1-g)^2 \mu^{11}_I + g^2 \mu^{22}_I + 2 g (1-g) \mu^{12}_I
\end{equation}
\begin{eqnarray}
M_D &=& (1-g)^2 \mu^{11}_D + g^2 \mu^{22}_D + 2 g (1-g) \mu^{12}_D \nonumber \\ &+& 2 g' l [ g (\mu^{22}_I + \mu^{22}_D) - (1-g) (\mu^{11}_I + \mu^{11}_D) + (1 - 2g) (\mu^{12}_I + \mu^{12}_D)] \nonumber \\ &+& (g' l)^2 [ \mu^{11}_I + \mu^{11}_D + \mu^{22}_I + \mu^{22}_D - 2 (\mu^{12}_I + \mu^{12}_D) ]
\label{MD}
\end{eqnarray}
the rotation tensor
\begin{equation}
\WW = \muO + \muT - 2 \muOT
\end{equation}
the translation-rotation coupling tensor
\begin{equation}
\GG = \Gamma_I \II + \Gamma_D \nn \nn
\end{equation}
with components
\begin{equation}
\Gamma_I = g \mu^{22}_I - (1-g) \mu^{11}_I  + (1-2g) \mu^{12}_I
\label{GI}
\end{equation}
\begin{eqnarray}
\Gamma_D &=& g \mu^{22}_D - (1-g) \mu^{11}_D  + (1-2g) \mu^{12}_D \nonumber \\ &+& g' l [  \mu^{11}_I + \mu^{11}_D + \mu^{22}_I + \mu^{22}_D - 2 (\mu^{12}_I + \mu^{12}_D) ]
\label{Ggen}
\end{eqnarray}
the translational phoretic mobility
\begin{equation}
\Vpho =   (1-g) \sgm_1 + g \sgm_2 + g' l \nn \nn (\sgm_2 -\sgm_1)
\end{equation}
and the internal (elongation-orientation) phoretic mobility
\begin{equation}
\vpho = \sgm_2 -\sgm_1 
\end{equation}

Importantly, we note that, while the internal phoretic mobility $\vpho$ and the rotation tensor $\WW$ are independent of the choice of tracking point as given by $g(l)$; the translational phoretic mobility $\Vpho$, the translation tensor $\MM$, and the translation-rotation coupling tensor $\GG$ do depend on this choice.

\subsection{Long-time diffusion-drift: Moment expansion} \label{moment}

We are interested in the long-time behaviour of the dumbbell enzyme. One can identify a hierarchy of timescales in which relaxation along the elongation coordinate $l$ is fastest, relaxation in orientation space $\nn$ is slower, and translational diffusion of the position coordinate $\RR$ is slowest \cite{illi17b,adel19a,adel19b}. The equations can then be pre-averaged assuming instantaneous equilibrium of the elongation $l$, and the orientation $\nn$ can be dealt with using an expansion in the moments of the orientation field \cite{adel19b}. Following such a procedure, one can show that the enzyme tends to align with gradients of the substrate \emph{via} phoresis, and with gradients of the concentration of the enzyme itself \emph{via} hydrodynamic interactions \cite{adel19b}. Moreover, one can write an evolution equation for the position probability distribution $\rho(\RR;t) = \int \mathrm{d}\LL \, \rho_{12}(\RR, \boldsymbol{l}; t)$ at long times, which reads
\begin{equation}
\partial_t \rho (\RR; t) = \nabla \cdot [ D_\mathrm{eff} \nabla \rho - \rho \Lambda_\mathrm{eff}  \nabla \rho_\mathrm{s} ]
\end{equation}
where, here and in the following, gradients and divergences are implied to be over position, i.e.~$\nabla \equiv \nabla_{\RR}$. The effective diffusion coefficient is given by
\begin{equation}
\frac{D_\mathrm{eff}}{\kB T} = \moy{M_I}+\frac{1}{3}\moy{M_D} - \frac{2}{3}\frac{\moy{\Gamma_I/l}^2}{\moy{W_I/l^2}} 
\label{Deff}
\end{equation}
which was also found in Refs.~\cite{illi17b,adel19a}, and the effective phoretic mobility is given by
\begin{equation}
\Lambda_\mathrm{eff} =  \moy{\Lambda^v_I} + \frac{1}{3} \moy{\Lambda^v_D} - \frac{2}{3} \frac{\moy{\Gamma_I/l}}{\moy{W_I/l^2}} \moy{\Lambda^\omega_I}
\label{PMeff}
\end{equation}
Here and in the following, the average $\moy{A}$ of any quantity $A$ is defined over a Bolzmann distribution of elongations, i.e. $\moy{A} \equiv N^{-1} \int_{a_1+a_2}^\infty \mathrm{d}l \, l^2 A \, \mathrm{e}^{-U(l)/\kB T}$ with normalization constant $N \equiv \int_{a_1+a_2}^\infty \mathrm{d}l \, l^2  \mathrm{e}^{-U(l)/\kB T}$, where the lower bound of the integrals arises due to the hard sphere interactions between the subunits, which limits the values of the elongation to $l>a_1+a_2$. The values of all these averages thus depend on the specific form of the potential $U(l)$ which holds the two subunits of the dumbbell together, e.g. on the rest length and on the rigidity or softness of this potential. We note that both the effective diffusion coefficient and the effective phoretic mobility have a similar structure, consisting of the first two terms, which correspond to the average of the contributions due to the translational modes, plus a third term which represents a fluctuation-induced correction \cite{illi17b,adel19a,adel19b}.

It is important to note that the pre-averaging and moment expansion procedures used to derive (\ref{Deff}) and (\ref{PMeff}) imply several approximations. Unfortunately, it is not possible to obtain an exact result for the long-time diffusion coefficient and phoretic mobility in the presence of an arbitrary substrate gradient $\rho_\mathrm{s}(\RR)$. However, we will show in Section~\ref{taylor} that in the absence of a gradient one can obtain an exact result for the long-time diffusion coefficient, which turns out to be similar but not exactly identical to the moment expansion result (\ref{Deff}).

\subsection{Binding-unbinding kinetics of the enzyme}\label{binding}

We have shown above that the long-time diffusion coefficient and phoretic mobility of a flexible dumbbell-like object depends on the form of the potential $U(l)$ holding the two subunits together. Moreover, the phoretic mobility will also depend on the surface properties of the subunits, in particular on the non-contact interactions $\phi^{is}$ between each subunit $i=1,2$ and the substrate molecules, which enter into the definition of the coefficients $A_i$ and $B_i$, see Eqs. (\ref{Aig}) and (\ref{Big}). However, enzymes do not interact with their substrate only through hydrodynamic and non-contact interactions: substrate molecules may also bind specifically to a binding pocket within the enzyme, a process which is typically accompanied by conformational changes of the enzyme itself. In the context of the dumbbell model, these conformational changes will involve changes in the potential $U(l)$ holding the two subunits together. Moreover, the surface properties of the subunits may also be affected by binding, in which case we would have different phoretic coefficients $A_i$ and $B_i$ in the free state and in the enzyme-substrate complex state.

As a consequence of these binding-induced changes, the effective diffusion coefficients of the free enzyme and the complex (denoted as $D_\mathrm{e}$ and $D_\mathrm{c}$), which are given by (\ref{Deff}) where the averages are taken using either the potential of the free enzyme $U_\mathrm{e}(l)$ or that of the complex $U_\mathrm{c}(l)$, will in general be different from each other, i.e.~$D_\mathrm{e}\neq D_\mathrm{c}$. The same is true for the phoretic mobilities of the free enzyme and the complex as given by (\ref{PMeff}), that is, we will have a different mobility for each state $\Lambda_\mathrm{e} \neq \Lambda_\mathrm{c}$. Including the specific binding and unbinding of the enzyme to the substrate thus requires us to consider the two coupled diffusion-drift equations \cite{agud18a}
\begin{eqnarray}
\partial_t \rho_\mathrm{e} (\RR; t) &=& \nabla \cdot [ D_\mathrm{e} \nabla \rho_\mathrm{e} - \rho_\mathrm{e} \Lambda_\mathrm{e}  \nabla \rho_\mathrm{s} ] - k_\mathrm{on} \rho_\mathrm{e} \rho_\mathrm{s} + k_\mathrm{off} \rho_\mathrm{c} \label{coup1} \\
\partial_t \rho_\mathrm{c} (\RR; t) &=& \nabla \cdot [ D_\mathrm{c} \nabla \rho_\mathrm{c} - \rho_\mathrm{c} \Lambda_\mathrm{c}  \nabla \rho_\mathrm{s} ] + k_\mathrm{on} \rho_\mathrm{e} \rho_\mathrm{s} - k_\mathrm{off} \rho_\mathrm{c} \label{coup2}
\end{eqnarray}
where $k_\mathrm{on}$ and $k_\mathrm{off}$ are the binding and unbinding rates of the substrate to the enzyme, respectively. In the limit in which binding-unbinding occurs much faster than the time it takes to diffuse into regions of space with significantly different substrate concentration, we can assume local equilibration with $k_\mathrm{on} \rho_\mathrm{e} \rho_\mathrm{s} \approx k_\mathrm{off} \rho_\mathrm{c}$ at every point. Under this assumption, one can obtain  an expression for the evolution of the total enzyme concentration (in both the free and complex states) $ \rho_\mathrm{e}^\mathrm{tot} \equiv  \rho_\mathrm{e} + \rho_\mathrm{c}$, which reads \cite{agud18a}
\begin{eqnarray}
\partial_t \rho_\mathrm{e}^\mathrm{tot} (\RR;t) = \nabla\cdot \left\{ D(\RR)\cdot\nabla \rho_\mathrm{e}^\mathrm{tot} - [\boldsymbol{V}_\mathrm{ph}(\RR)+\boldsymbol{V}_\mathrm{bi}(\RR)] \rho_\mathrm{e}^\mathrm{tot} \right\}.
\label{eq:totalevol}
\end{eqnarray}
with the substrate-dependent diffusion coefficient
\begin{eqnarray}
D(\RR) = D_\mathrm{e} + (D_\mathrm{c} - D_\mathrm{e}) \frac{\rho_\mathrm{s}(\RR)}{K + \rho_\mathrm{s}(\RR)}
\label{eq:Defff}
\end{eqnarray}
the substrate-dependent phoretic drift velocity
\begin{eqnarray}
\boldsymbol{V}_\mathrm{ph}(\RR) = \left[\Lambda_\mathrm{e} + (\Lambda_\mathrm{c} - \Lambda_\mathrm{e}) \frac{\rho_\mathrm{s}(\RR)}{K + \rho_\mathrm{s}(\RR)}\right] \nabla \rho_\mathrm{s}
\label{eq:Vph}
\end{eqnarray}
and the binding-induced drift velocity
\begin{eqnarray}
\boldsymbol{V}_\mathrm{bi}(\RR) =  -(D_\mathrm{c} - D_\mathrm{e}) \frac{K}{[K + \rho_\mathrm{s}(\RR)]^2} \nabla \rho_\mathrm{s}(\RR).
\label{eq:Vbi}
\end{eqnarray}
where $K\equiv k_\mathrm{off}/k_\mathrm{on}$ is the dissociation constant of the substrate.

With increasing substrate concentration $\rho_\mathrm{s}$, both the diffusion coefficient (\ref{eq:Defff}) as well as the phoretic velocity (\ref{eq:Vph}) vary smoothly, with a Michaelis-Menten-type dependence, from the value corresponding to the free state in the absence of substrate ($\rho_\mathrm{s}=0$) to the value for the complex state at saturating substrate concentration ($\rho_\mathrm{s} \gg K$). The binding-induced velocity (\ref{eq:Vbi}) points towards decreasing substrate concentrations in the case of binding-induced enhanced diffusion ($D_\mathrm{c} > D_\mathrm{e}$), and towards increasing substrate concentrations for inhibited diffusion ($D_\mathrm{c} < D_\mathrm{e}$). As will be shown in Section~\ref{steady}, the phoretic and binding-induced velocities can pull the enzyme in opposing directions, leading to non-trivial steady state profiles for the enzyme, which may accumulate in or be depleted from regions with a specific substrate concentration.

\section{Long-time diffusion coefficient in the absence of a gradient: Generalized Taylor dispersion theory}\label{taylor}

\subsection{General expression for the diffusion coefficient}

In Section~\ref{moment}, we discussed the long time diffusion coefficient for a flexible dumbbell as calculated using an approximate method, which involves pre-averaging over the elongation degree of freedom $l$ of the enzyme, and using a moment expansion scheme to deal with the orientation degree of freedom $\nn$ \cite{illi17b,adel19a,adel19b}. However, in the absence of a substrate gradient, it is also possible to obtain the exact long-time diffusion coefficient without recurring to any approximations by means of generalized Taylor dispersion theory (GTDT) \cite{bren82,bren87,bren93}. GTDT applies to systems in which the phase space is divided into two orthogonal subspaces consisiting of local and global coordinates, which are bounded and unbounded respectively: in our particular case, the local coordinate is $\LL$, which includes both orientation and elongation, while the global coordinate is the position $\RR$. Particularized to the evolution equation (\ref{smol1}) for the dumbbell probability distribution, in the absence of a gradient $\nabla \rho_\mathrm{s}=0$, GTDT implies the long time diffusion coefficient\footnote{In particular, see Eq.~(5.13) in Ref.~\cite{bren87}.}
\begin{equation}
\frac{D_\mathrm{eff}}{\kB T} = N'^{-1} \int \dd \LL \, \ex{-U(l)/\kB T} (\MM - \GG \cdot \WW^{-1} \cdot \GG)
\end{equation}
where $N'\equiv \int \dd \LL \ex{-U(l)/\kB T}$ is a normalization constant. Introducing the expressions for $\MM$, $\GG$, and $\WW$ above, and performing the integral over orientations, we finally obtain the result
\begin{equation}
\frac{D_\mathrm{eff}}{\kB T} =  \moy{M_I} + \frac{1}{3} \moy{M_D}  - \frac{2}{3} \moy{\frac{(\Gamma_I)^2}{W_I}} - \frac{1}{3} \moy{\frac{(\Gamma_I + \Gamma_D)^2}{W_I + W_D}}
\label{deff1}
\end{equation}
This result bears strong similarities to the result obtained using moment expansion (\ref{Deff}). In particular, both results include the first two terms, which correspond to the average of the contributions due to the translational modes $M_I + \frac{1}{3} M_D$, plus negative fluctuation-induced corrections \cite{illi17b,adel19a}.  However, the two results also show some key differences: (i) the averaging structure of the third term is different (the elongation $l$ does not appear explicitly in the GTDT result, and we take an average of the square rather than a square of the average, which further highlights the fluctuation-induced origin of the corrections), and (ii) there is an extra term in the GTDT result, which goes with $\Gamma_I + \Gamma_D$. These differences can be traced back to the use of a pre-averaging approximation over the elongation $l$ in the moment expansion method, and have some important consequences, as described below.

\subsection{Independence of the choice of tracking point and non-existence of a `center of diffusion'}

Firstly, we can substitute the explicit expressions (\ref{Mgen}--\ref{Ggen}) for the mobility tensors as a function of the particular choice of tracking point $g(l)$ into expressions (\ref{Deff}) and (\ref{deff1}) for the long time diffusion coefficient. It is straightforward to check that the approximate expression (\ref{Deff}) still depends on $g(l)$. One would however expect that, in the long time limit, the diffusion of a finite sized object such as the dumbbell under consideration would become independent of the choice of tracking point. This is indeed true for the exact expression (\ref{deff1}), for which the dependence on $g(l)$ cancels out, proving that indeed the long time diffusion coefficient is independent of the choice of tracking point. This implies that, while at short times the enzyme will diffuse with a diffusion coefficient $\kB T [ \moy{M_I} + \frac{1}{3} \moy{M_D}]$ which depends on the choice of tracking point $g(l)$, at long times this diffusion coefficient will decrease and cross over into a choice-independent diffusion coefficient given by (\ref{deff1}).

One may wonder whether there is a particular choice of tracking point which would diffuse at all times with a time-independent diffusion coefficient as given by (\ref{deff1}), i.e.~a `center of diffusion' \cite{wege85,cich19}. This would be possible only if the two correction terms in (\ref{deff1}) were identically zero for some choice of $g(l)$. However, because each of the two terms is the average of a nonnegative quantity, the total correction can only be zero if both $\Gamma_I = 0$ and $\Gamma_D = 0$. Noting the form of $\Gamma_I$ and $\Gamma_D$ in  (\ref{GI}--\ref{Ggen}), we see that such a choice of $g(l)$ does not exist in general, and thus no center of diffusion exists in general for such a flexible object. A center of diffusion does exist in the particular case in which the dumbbell is symmetric, with $\muO=\muT$, in which case we have $\Gamma_I = \Gamma_D = 0$ if the tracking point is chosen as the midpoint between subunits, i.e. $g(l)=1/2$. A second particular (limiting) case for which a center of diffusion exists is that of a very long dumbbell for which hydrodynamic interactions between subunits become negligible, i.e.~when we have $\muOT=0$ and $\boldsymbol{\mu}^{ii} = (6\pi \eta a_i)^{-1} \II$. In this case, the location of the center of diffusion is given by
\begin{equation}
g(l) =  \frac{1}{2} + \frac{1}{2} \frac{\mu^{11} - \mu^{22}}{ \mu^{11} + \mu^{22}} = \frac{1}{2} + \frac{1}{2} \frac{a_2 - a_1}{a_1+a_2}
\label{dg}
\end{equation}

\subsection{Explicit expansion for the diffusion coefficient}

Secondly, we can use the mobility tensors in the Oseen approximation
\begin{equation}
\boldsymbol{\mu}^{ii} = \frac{1}{6 \pi \eta a_i} \II + O(a^3/\eta l^4)
\end{equation}
\begin{equation}
\muOT =  \frac{1}{8\pi \eta l}\left(\II+\nn\nn\right)  + O(a^2/\eta l^3)
\end{equation}
and introduce them into (\ref{deff1}). Expanding everything in powers of $1/l$, we can write the long-time diffusion coefficient as
\begin{equation}
\frac{D_\mathrm{eff}}{\kB T} = \frac{1}{6 \pi \eta (a_1 + a_2)} \left[ 1 + 2 \frac{a_1 a_2}{a_1+a_2} \moy{\frac{1}{l}} - \frac{9}{8} \frac{a_1 a_2 (a_1-a_2)^2}{(a_1+a_2)^2}  \moy{\frac{1}{l^2}} + O\left( \moy{\frac{a_i^3}{l^3}} \right) \right]
\label{deff3}
\end{equation}
Note that obtaining the third order would require us to go beyond the Oseen approximation of $\muOT$, and going to fourth order would also require the higher order terms in $\boldsymbol{\mu}^{ii}$ but, conceptually, obtaining the higher order terms is a straightforward procedure. 

The expression (\ref{deff3}) for the long time diffusion coefficient is particularly simple and transparent, and could not have been obtained from the approximate result (\ref{Deff}), due to the different averaging structure and the missing last term. One key feature of this result is that the leading order of the long-time diffusion coefficient, i.e.~the diffusion coefficient of a very long dumbell for which hydrodynamic interactions are negligible, goes as $\frac{\kB T}{6 \pi \eta (a_1 + a_2)}$. This leading order contribution is independent of the particular kind of potential holding the two subunits together (or how stiff of soft this potential is) and of the hydrodynamic interactions between subunits, and, most importantly, it is independent of the length of the dumbell. In particular, the leading order \emph{does not} go as $\sim \frac{\kB T}{\eta l}$, as one may have expected, see e.g.~\cite{vish19}. The leading order contribution also shows that, in the limit of one of the subunits being much larger than the other, e.g.~$a_1 \gg a_2$, the diffusion coefficient of the dumbbell tends to the diffusion coefficient of the larger subunit $\frac{\kB T}{6 \pi \eta a_1}$.

Beyond the leading order, we find a first order term in $\moy{1/l}$, i.e. related to the average shape of the dumbbell, which is always positive. More interesting is the second order term going as $\moy{1/l^2}$, which includes the contribution due to the thermal fluctuations about the average shape of the dumbbell. This fluctuation-induced correction is always negative, and becomes zero only in the particular case of a symmetric dumbbell with $a_1=a_2$, as found previously \cite{illi17b,adel19a}.

\subsection{Changes in diffusion due to substrate binding}

In Section~\ref{binding}, we mentioned how specific binding of the enzyme to a substrate molecule is expected to modify the diffusion coefficient of the enzyme, through a modification of the potential $U(l)$ which holds the two subunits together. Using the result (\ref{deff3}) for the long time diffusion coefficient, we can now explicitly discuss these changes. Assuming that the potentials in the free enzyme state and enzyme-substrate complex state are given by $U_\mathrm{e}(l)$ and $U_\mathrm{c}(l)$, respectively, we can calculate the two diffusion coefficients $D_\mathrm{e}$ and $D_\mathrm{c}$ by taking the averages in (\ref{deff3}) using the corresponding potential. The relative change in diffusion coefficient between the free and complex states can then be calculated as
\begin{eqnarray}
\alpha \equiv \frac{D_\mathrm{c} - D_\mathrm{e}}{D_\mathrm{e}} =&&  2 \frac{a_1 a_2}{a_1+a_2} \left( 1 - 2 \frac{a_1 a_2}{a_1+a_2} \moy{\frac{1}{l}}_\mathrm{e} \right) \left( \moy{\frac{1}{l}}_\mathrm{c} - \moy{\frac{1}{l}}_\mathrm{e}  \right) \nonumber \\ &&  - \frac{9}{8} \frac{a_1 a_2 (a_1-a_2)^2}{(a_1+a_2)^2} \left(  \moy{\frac{1}{l^2}}_\mathrm{c} - \moy{\frac{1}{l^2}}_\mathrm{e} \right) + O \left( \frac{a_i^3}{l^3} \right)
\label{deff4}
\end{eqnarray}

To lowest order, this diffusion change is governed by the changes in the average length of the dumbbell, with $\alpha \approx 2 \frac{a_1 a_2}{a_1+a_2} \left( \moy{\frac{1}{l}}_\mathrm{c} - \moy{\frac{1}{l}}_\mathrm{e}  \right)$. This implies that the complex will diffuse faster than the enzyme, and thus the presence of the substrate will lead to enhanced diffusion, if the average length of the dumbbell is shorter in the complex state than in the free state, i.e.~if $\moy{\frac{1}{l}}_\mathrm{c} > \moy{\frac{1}{l}}_\mathrm{e}$. This will be the case either if (i) the preferred length of the potential is shortened, or (ii) the potential becomes more stiff, in which case the entropic tendency of thermal fluctuations to stretch the dumbbell is more strongly counteracted. This can be checked explicitly for a harmonic potential $U_i(l) = k_i (l - l_i)^2/2$ where $l_i$ is the preferred length, $k_i$ is the stiffness, and the subindex $i=\mathrm{e,c}$  represents the free or complex state of the enzyme. For a sufficiently stiff potential, we can neglect the possibility of direct subunit-subunit contact and thus ignore the lower bound in thermal averages $\moy{A}$ of any quantity $A$, so that $\int_{a_1+a_2}^\infty \mathrm{d}l \, l^2 A \, \mathrm{e}^{-U(l)/\kB T} \approx \int_{-\infty}^\infty \mathrm{d}l \, l^2 A \, \mathrm{e}^{-U(l)/\kB T}$. In this way, we obtain the simple expression
\begin{equation}
\moy{\frac{1}{l}}_i \approx \frac{1}{l_i} \frac{k_i}{\frac{\kB T}{l_i^2} + k_i}
\label{avgl}
\end{equation}
which shows explicitly that increasing the stiffness $k_i$ increases the average inverse length $\moy{\frac{1}{l}}_i$ and thus increases the diffusion coefficient.

The next order, in particular the term in the second line of (\ref{deff4}) proportional to $\left(  \moy{\frac{1}{l^2}}_\mathrm{c} - \moy{\frac{1}{l^2}}_\mathrm{e} \right)$, includes the contributions due to changes in the fluctuations of the dumbbell length. Thus, even if the average length of the dumbbell remains unchanged, changes in fluctuations are sufficient to induce a change in the diffusion coefficient. If we denote the variance of the inverse dumbbell length as $s_i^2 \equiv \moy{\frac{1}{l^2}}_i - \moy{\frac{1}{l}}_i^2$, which is a measure of the strength of the fluctuations, setting $\moy{\frac{1}{l}}_\mathrm{c} = \moy{\frac{1}{l}}_\mathrm{e}$ in (\ref{deff4}) shows that the change in diffusion coefficient  purely due  to changes in the strength of fluctuations will go as $\alpha \approx  - \frac{9}{8} \frac{a_1 a_2 (a_1-a_2)^2}{(a_1+a_2)^2} \left( s^2_\mathrm{c} - s^2_\mathrm{e} \right)$. Therefore, making the enzyme more rigid ($ s^2_\mathrm{c} < s^2_\mathrm{e}$) will increase the diffusion coefficient, even if the average length remains unchanged.

The closed form expression (\ref{deff4}) for the diffusion enhancement due to substrate-induced rigidification of the enzyme agrees well with the results in Ref.~\cite{illi17b}. In that work, no such simple closed form expression could be found, due to the use of the approximate moment expansion scheme [see Eq.~(\ref{Deff})] rather than the exact result from generalized Taylor dispersion theory [see Eq.~(\ref{deff1})]. We note also that the prediction of enhanced diffusion due to rigidification of the enzyme has been quantitatively confirmed using Brownian dynamics simulations with hydrodynamic interactions in Ref.~\cite{kond19}, although it was argued in the latter work that the changes in average length of the enzyme arising from rigidification might be too large to be biologically relevant.

\section{Steady-state enzyme distribution in the presence of a gradient} \label{steady}

As argued in Section~\ref{binding} [see Eq.~(\ref{eq:totalevol})], in the presence of an arbitrary substrate gradient, it is expected that an enzyme will undergo diffusion with a substrate-dependent diffusion coefficient, and moreover will experience a drift in the direction of the gradient which arises from two distinct contributions: a phoretic contribution due to non-contact interactions between the enzyme subunits and the substrate molecules, and a binding-induced contribution due to conformational changes of the enzyme when it binds to a substrate molecule to form a complex.

In Ref.~\cite{agud18a} we described how, for a typical enzyme, we expect the non-contact interactions (van der Waals, electrostatic...) to be attractive, while specific binding of the substrate usually leads to enhanced diffusion ($D_\mathrm{c}>D_\mathrm{e}$). As a consequence, the two contributions to the drift velocity point in opposite directions (the phoretic contribution towards the substrate, the binding-induced one away from the substrate) and compete against each other. This competition between two contributions to chemotaxis can explain \cite{agud18a} the conflicting experimental observations regarding whether urease chemotaxes towards \cite{seng13} or away from \cite{jee17} urea. For simplicity, let us focus on the case in which the phoretic mobility of the enzyme is unchanged by binding, i.e.~we have $\Lambda_\mathrm{e} = \Lambda_\mathrm{c} = \Lambda$ and thus, \emph{via} Eq.~(\ref{eq:Vph}), we have $\boldsymbol{V}_\mathrm{ph}(\RR) = \Lambda \nabla \rho_\mathrm{s}$. The binding-induced velocity, on the other hand, is given by Eq.~(\ref{eq:Vbi}). Because the binding-induced velocity becomes weaker with increasing substrate concentration, whereas the phoretic velocity is independent of substrate concentration, imposing  $|\boldsymbol{V}_\mathrm{ph}|=|\boldsymbol{V}_\mathrm{bi}|$ we find a critical substrate concentration
\begin{eqnarray}
\rho_\mathrm{s}^* \equiv K \left( \sqrt{ \frac{D_\mathrm{e} |\alpha|}{|\Lambda| K} } - 1 \right)
\label{eq:critical}
\end{eqnarray}
above and below which phoresis and binding-induced enhanced diffusion dominate, respectively. As above, $\alpha \equiv (D_\mathrm{c} - D_\mathrm{e})/D_\mathrm{e}$  represents the relative change in diffusion coefficient between the free and complex states of the enzyme.

More generally, besides a competition between attractive phoresis and enhanced diffusion, we could also imagine nanoscale objects (enzymes or otherwise) which experience a competition between repulsive phoresis and inhibited diffusion. In the former case, the enzyme will move towards higher substrate concentrations in regions with $\rho_\mathrm{s}>\rho_\mathrm{s}^*$ (attractive phoresis dominates), or toward lower substrate concentration in regions with $\rho_\mathrm{s}<\rho_\mathrm{s}^*$ (enhanced diffusion dominates), so that it will be depleted from regions with the critical concentration $\rho_\mathrm{s}^*$. In the latter case, the enzyme will move towards lower substrate concentrations in regions with $\rho_\mathrm{s}>\rho_\mathrm{s}^*$ (repulsive phoresis dominates), or toward higher substrate concentration in regions with $\rho_\mathrm{s}<\rho_\mathrm{s}^*$ (inhibited diffusion dominates), so that it will accumulate in regions with the critical concentration $\rho_\mathrm{s}^*$.

We will now demonstrate that the dynamical argument summarized in the previous paragraph is also reflected in the long time steady-state distribution of enzyme or enzyme-like nano-objects $\rho_\mathrm{e}^\mathrm{tot} (\RR)$ in the presence of a sustained spatial gradient of substrate $\rho_\mathrm{s}(\RR)$. In fact, we can directly calculate this steady-state enzyme distribution from the evolution equation (\ref{eq:totalevol}). We note that the evolution equation has the form $\partial_t \rho_\mathrm{e}^\mathrm{tot} = - \nabla \cdot \boldsymbol{J}_\mathrm{e}^\mathrm{tot}$, where $\boldsymbol{J}_\mathrm{e}^\mathrm{tot}$ is the total flux of enzyme. Assuming that there are no sources or sinks of enzyme, we can set $\boldsymbol{J}_\mathrm{e}^\mathrm{tot} = 0$ to obtain the steady-state distribution
\begin{equation}
\rho_\mathrm{e}^\mathrm{tot}(\RR) = \rho_0 \frac{\exp \left(\frac{\Lambda \rho_\mathrm{s}}{D_\mathrm{e}(1+\alpha)} \right) \left[1 + (1+\alpha)\frac{\rho_\mathrm{s}}{K} \right]^{\frac{\Lambda K \alpha}{D_\mathrm{e}(1+\alpha)^2}}}{1+\alpha \frac{\rho_\mathrm{s}}{K+\rho_\mathrm{s}}}
\label{steady1}
\end{equation}
where $\rho_0$ is a normalization constant (corresponding to the enzyme concentration at points where $\rho_\mathrm{s}=0$), which can be used to enforce a constraint on the total number of enzymes in solution.

We note that, in the limit in which phoresis is negligible $\Lambda \to 0$, or equivalently when the substrate concentration is very small $\rho_\mathrm{s} \ll \rho_\mathrm{s}^*$, the steady state distribution becomes $\rho_\mathrm{e}^\mathrm{tot}(\RR) = \rho_0  \left( 1+\alpha \frac{\rho_\mathrm{s}}{K+\rho_\mathrm{s}} \right)^{-1}$, i.e.~it is inversely proportional to the diffusion coefficient (\ref{eq:Defff}), with $\rho_\mathrm{e}^\mathrm{tot}(\RR) \propto 1/D(\RR)$, as experimentally observed in Ref.~\cite{jee17}. This reflects the fact that, in the absence of phoresis, the evolution equation (\ref{eq:totalevol}) can be written as $\partial_t \rho_\mathrm{e}^\mathrm{tot} (\RR;t) = \nabla^2 \left\{ D(\RR) \rho_\mathrm{e}^\mathrm{tot} \right\}$. In the opposite limit in which binding-induced changes in diffusion are negligible, given by $\alpha \to 0$ or equivalently for high substrate concentrations $\rho_\mathrm{s} \gg \rho_\mathrm{s}^*$, the steady state distribution becomes $\rho_\mathrm{e}^\mathrm{tot}(\RR) = \rho_0 \exp \left( \frac{\Lambda \rho_\mathrm{s}}{D_\mathrm{e}} \right)$.

Using the definition (\ref{eq:critical}) of the critical substrate concentration $\rho_\mathrm{s}^*$, the steady-state distribution (\ref{steady1}) can alternatively be written in the form
\begin{equation}
\rho_\mathrm{e}^\mathrm{tot}(\RR) = \rho_0  \frac{\exp \left(\frac{\alpha}{1+\alpha} \frac{\rho_\mathrm{s} K}{(K+\rho_\mathrm{s}^*)^2} \right) \left[1 + (1+\alpha)\frac{\rho_\mathrm{s}}{K} \right]^{\left(\frac{\alpha}{1+\alpha}\frac{K}{K+\rho_\mathrm{s}^*}\right)^2}}{1+\alpha \frac{\rho_\mathrm{s}}{K+\rho_\mathrm{s}}}
\label{steady2}
\end{equation}
which is valid only in cases in which phoresis and binding-induced changes in diffusion compete against each other, i.e. when $\Lambda$ and $\alpha$ are either both positive or both negative, but has the advantage of making the relation to the critical concentration $\rho_\mathrm{s}^*$ more explicit.

\begin{figure}
\centering
\includegraphics[width=1\linewidth]{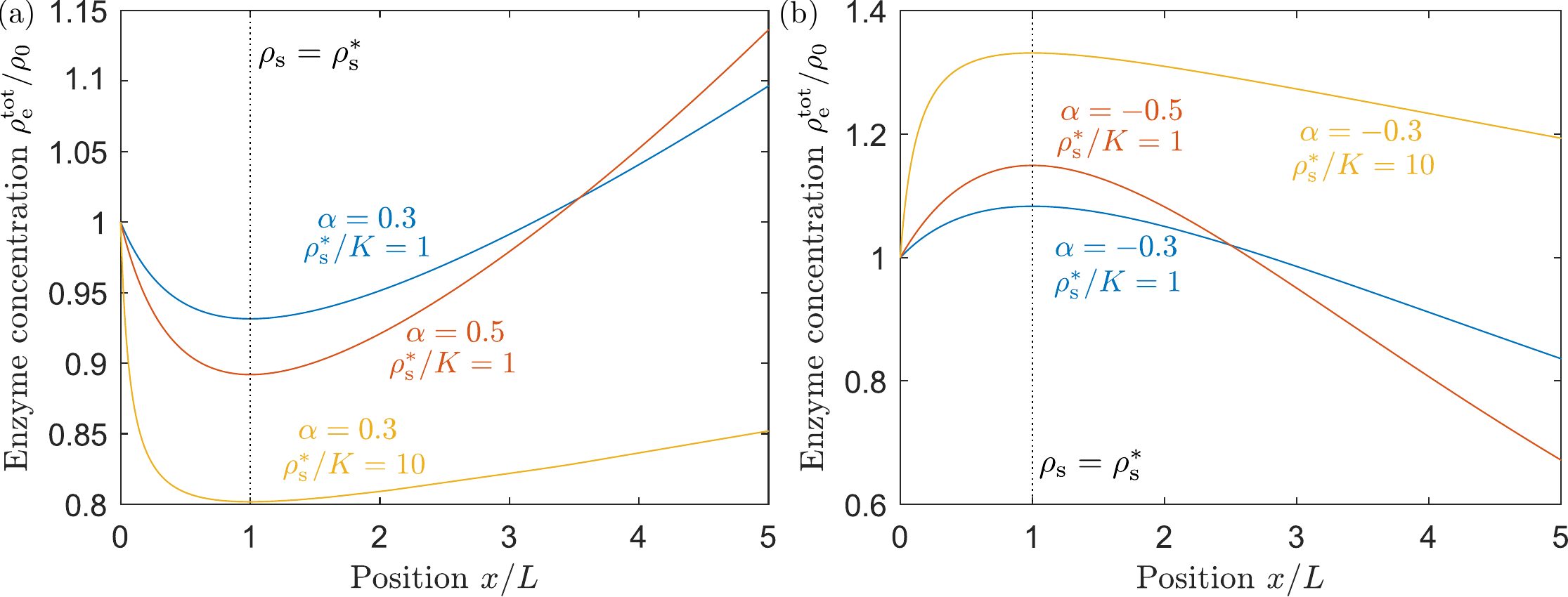}
\caption{Steady state concentration of enzyme (or enzyme-like nano-object), as given by (\ref{steady1}--\ref{steady2}), in the presence of a sustained linear gradient of substrate which reaches the critical concentration $\rho_\mathrm{s}^*$ at position $x=L$. In (a), the enzyme exhibits enhanced diffusion and attractive phoresis, and is repelled from the point $x=L$. In (b), the enzyme exhibits inhibited diffusion and repulsive phoresis and is attracted to the point $x=L$.}
   \label{fig:steady}
\end{figure}

In Figure~\ref{fig:steady}, we plot the steady state distribution of enzyme $\rho_\mathrm{e}^\mathrm{tot}(x)$ in the presence of a one-dimensional linear gradient of substrate given by $\rho_\mathrm{s}(x) = \rho_\mathrm{s}^* x/L $, which would correspond to a channel of length $nL$ with a substrate sink with $\rho_\mathrm{s}=0$ at one end ($x=0$) and a substrate reservoir with concentration $\rho_\mathrm{s}=n \rho_\mathrm{s}^*$ at the other end ($x=nL$). In this case, the substrate concentration at $x=L$ is exactly the critical concentration, i.e.~$\rho_\mathrm{s}(L) =\rho_\mathrm{s}^*$. We find that, as predicted, the enzyme is depleted from the point $x=L$ in the case with enhanced diffusion and attractive phoresis ($\alpha,\Lambda>0$), see Figure~\ref{fig:steady}(a), but accumulates at $x=L$ for inhibited diffusion and repulsive phoresis ($\alpha,\Lambda<0$), see Figure~\ref{fig:steady}(b). The strength of the depletion/accumulation effect increases both with the magnitude of $\alpha$ and with $\rho_\mathrm{s}^*$, with the enzyme concentration reaching minimal and maximal values of $\rho_\mathrm{e}^\mathrm{tot}/\rho_0 \approx 1/(1+\alpha)$ at $x=L$ in the limit of $\rho_\mathrm{s}^* \gg K$.

It should be noted that, although the enzyme concentration profiles given by (\ref{steady1}--\ref{steady2}) are steady-state profiles corresponding to a zero-flux condition $\boldsymbol{J}_\mathrm{e}^\mathrm{tot} = 0$, they do not correspond to an equilibrium Boltzmann distribution. In fact, the enzyme profiles are a consequence of the externally imposed substrate gradient, and are therefore due to intrinsically non-equilibrium effects. If the substrate gradient is not artificially sustained, the system will tend to an equilibrium distribution with uniform substrate and enzyme concentrations. In relation to this, it should be noted that typical enzyme chemotaxis experiments \cite{seng13,seng14,dey14,yu09,zhao17} are not performed in a sustained non-equilibrium steady-state, but rather measure the early stages of the transient dynamics from an initial state where enzyme and substrate are not fully mixed towards the uniform equilibrium distribution. Detailed theoretical modelling of such experiments therefore requires the solution of the time evolution equation (\ref{eq:totalevol}), or more generally (if binding-unbinding cannot be considered to be sufficiently fast) of the coupled equations (\ref{coup1}--\ref{coup2}), with appropriate initial and boundary conditions.

\section{Conclusion}

As mentioned in the introduction, the mechanisms underlying enhanced diffusion and chemotaxis of enzymes in the presence of their substrate are still far from being understood. In fact, it is likely that not just one but several different mechanisms are responsible for the observed behaviours. Each of these mechanisms may be more or less relevant for any given enzyme, and several of them may act simultaneously and add up to the overall enhanced diffusion or chemotaxis observed.

Besides the active mechanisms already discussed in the introduction, and the passive mechanisms that we have described in detail here, an intriguing possibility that has been recently proposed and investigated \cite{gunt18,zhan18,jee18,jee19} is that of substrate-induced dissociation of the enzymes. Indeed, many of the enzymes for which enhanced diffusion and chemotaxis has been observed are oligomeric enzymes, composed of monomeric subunits which may reversibly associate and dissociate. Careful experiments will thus be needed to discriminate between all these different mechanisms, and to identify which ones are relevant for any given enzyme under given experimental conditions.

Careful theoretical work will also be needed to characterize all the active and passive mechanisms that can contribute to enhanced diffusion and chemotaxis of enzymes. As shown here, passive mechanisms associated to binding-induced conformational changes as well as phoresis of the enzyme can provide an important contribution. Moreover, the results described here should be useful beyond understanding biological enzymes. In particular, the competition between phoretic effects and conformational changes that we have described (see Fig.~\ref{fig:steady}) may be harnessed in the design of synthetic nano-vehicles that are directed towards finely-tuned regions in space as determined by the specific concentration of a certain chemical.

\begin{acknowledgement}
We thank Tunrayo Adeleke-Larodo and Pierre Illien for stimulating collaborations on enzyme-related topics. We acknowledge funding from the U.S. National Science Foundation under MRSEC Grant No. DMR-1420620.
\end{acknowledgement}

\bibliographystyle{ieeetr}
\bibliography{biblio}

%
%

\end{document}